 \documentclass[final,3p]{elsarticle}

\usepackage{graphicx,subcaption,url}
\usepackage{float}

\usepackage{amsmath,amsthm,amssymb}

\biboptions{sort&compress} 

\newcounter{bla}

\newcommand{\be}{\begin{equation}}
\newcommand{\ee}{\end{equation}}
\newcommand{\ba}{\begin{eqnarray}}
\newcommand{\ea}{\end{eqnarray}}

\def\lsim{\raise0.3ex\hbox{$\;<$\kern-0.75em\raise-1.1ex\hbox{$m\;$}}}
\def\gsim{\raise0.3ex\hbox{$\;>$\kern-0.75em\raise-1.1ex\hbox{$m\;$}}}
\def\eps{\varepsilon}
\def\theta{\vartheta}

\def\d{{\rm d}}
\def\e{{\rm e}}

\usepackage{xcolor,color}
\definecolor{darkBlue}{rgb}{0, 0, 0.8}

\def\OBA{OB~association}

\renewcommand{\vec}[1]{\boldsymbol{#1}}

\newcommand{\aap}{{Astron.\ Astrophys. }}
\newcommand{\aaps}{{Astron.\ Astrophys. Suppl.}}
\newcommand{\aj}{{Astron.\ J. }}
\newcommand{\apj}{{Astrophys.\ J. }}
\newcommand{\apjl}{{Astrophys.\ J.\ Lett. }}
\newcommand{\apjs}{{Astrophys.\ J.\ Suppl. }}

\newcommand{\araa}{{Ann.\ Rev.\  Astron.\ Astrophys. }}

\newcommand{\mnras}{{Mon.\ Not.\ Roy.\ Astron.\ Soc. }}
\newcommand{\nar}{{New Astron.\ Rev. }}

\journal{Computer Physics Communications}

\begin{document}

\begin{frontmatter}



\title{Galactic Distribution of Supernovae and OB Associations}


\author{M.~Kachelrie{\ss} and V.~Mikalsen}

\address{Institutt for fysikk, NTNU, Trondheim, Norway}

\begin{abstract}
We update and extend a previous model by Higdon and Lingenfelter for the
longitudinal profile of the N\,II  intensity in the Galactic plane. The model
is based on four
logarithmic spiral arms, to which features like the Local Arm and local
sources are added. Connecting then the N\,II to the H\,II emission, we
use this model to determine the average spatial distribution of
OB~associations in the Milky Way. Combined with a stellar mass and 
cluster distribution function,
the model predicts the average spatial and temporal distribution of
core-collapse supernovae in the Milky Way. In addition to this average
population, we account for supernovae from observed OB~associations,
providing thereby a more accurate description of the nearby Galaxy.
The complete model is made publicly available in the python code {\tt SNOB}. 
\end{abstract}


\end{frontmatter}
%
%
{\bf PROGRAM SUMMARY}\\
%
\begin{small}
\noindent
{\em Manuscript Title:}
Galactic Distribution of Supernovae and OB Associations\\
{\em Program Title:} {\tt SNOB\,1.1}:
Simulating the distribution of SuperNovae and OB associations in the Milky Way
\\
{\em Journal Reference:}                                      \\
{\em Catalogue identifier:}                                   \\
{\em Licensing provisions:}                                   
CC by NC 3.0. 
\\
{\em Programming language:}  Python 3.8                     \\
{\em Computer:}                                               
Any computer with python version 3 installed                  \\ 
{\em Operating system:}  Any system with python version 3                \\
{\em RAM:} Depends on the parameter {\tt num\_grad\_subdivision}                                              
{\em Nature of problem:}
Determination of the distribution of OB~associations from the observed N\,II
line intensity; derivation of the resulting distribution of core-collapse
supernovae. 
   \\
{\em Solution method:}
Numerical integration of line-of-sight integrals for the N\,II line intensity;
Monte Carlo simulation of the spatial and time distribution of OB~associations
and core-collapse supernovae in the Milky Way.
\end{small}

\newpage
\tableofcontents

\section{Introduction}

The mechanical energy injected by supernovae (SNe) explosions into
the interstellar medium (ISM) has been a prime candidate to power the
acceleration of Galactic cosmic rays (CR) since the first suggestion by
Baade and Zwicky~\cite{1934PNAS...20..259B,1969ocr..book.....G,Hillas:2005cs}. 
In addition, these explosions generate the Galactic populations of stellar
black holes and pulsars, are the main source of heavier elements
up to the iron group, and drive the turbulence of the ISM. Thus SNe are
key factors for the evolution and dynamics of galaxies, and therefore
of high interest. Around 80\%--90\% of all SN explosions are core-collapse
SNe, marking the end of stellar fusion in stars with initial mass above
$\simeq 8\,M_{\odot}$~\cite{Woosley__weaver_1995, Smartt_2009}. Therefore, 
the progenitors of these stars are of the OB type, which are the main
emitters of ionizing radiation 
in the Milky Way. Moreover, the life-time of these massive stars
is short, and thus core-collapse SNe should happen close to their birthplace
in associations of OB stars.

An important ingredient in simulations of CR propagation and the prediction
of the primary and secondary CR fluxes is the source distribution of CR
sources. The details of the CRs source distribution become more important
for those CRs whose propagation distance in the Galaxy is short: Important
examples are electrons in the TeV energy range and above because of their
severe energy losses, and CR protons close to PeV energies because of
their fast escape from the Galaxy. For instance, it has been argued that a
significant change in the observed positron flux at Earth is caused by
the spiral structure of the Milky Way~\cite{Shaviv:2009bu}. Moreover,
the interpretation of the fluxes of radioactive cosmic ray isotopes is
significantly affected by how the effect of the Milky Way spiral arms is
included~\cite{DeLaTorreLuque:2024wtv}. In practise,
however, most often the CR source distribution is modelled using a
template based on the distribution of pulsars.
Since pulsars are normally not observed soon after their birth and
receive rather large kick velocities at birth, their spatial distribution
does not resolve the spiral structure of the Milky Way. Moreover, the rather
large errors of their distance estimates based on dispersion measures
have allowed in the past only the determination of the radial
distribution of pulsars~\cite{Green:2015isa,Ranasinghe:2022ntj,Xie_2024}.
Note, however, that accurate pulsar distances based on their parallaxes
may change this in the future.

As an alternative, Higdon and Lingenfelter suggested to use the distribution
of \OBA s as a proxy for the distribution Galactic CR
sources~\cite{Higdon_lingenfelter_2005,higdon_galactic_2013}. The use of such
a proxy is well justified for a rather generic class of CR sources, including
SN remnants, young pulsars, pulsar wind nebula,
and superbubbles or a combination of these source types. Moreover, the
position of \OBA s can be traced by the observation of their line emission,
choosing e.g.\ the H$\alpha$ or low-frequency radio recombination lines.
It is, however, more convenient to use instead the N\,II emission line, as in
this case absorption can be neglected. In particular, Higdon and Lingenfelter
suggested to use the lateral distribution of the N\,II $205\mu$m line
intensity observed by the FIRAS instrument on-board the COBE satellite.
In order to construct the three-dimensional distribution of \OBA s
from the two-dimensional sky picture of the N\,II $205\mu$m
intensity, a model for the spatial structure of the Milky Way
has to be supplemented which then can be fitted to the observational data.

The aim of the present work is to provide a simulation tool which generates
the spatial and temporal distribution of core-collapse supernovae in the
Milky Way. In this work, we follow closely the approach of Higdon and
Lingenfelter~\cite{Higdon_lingenfelter_2005,higdon_galactic_2013}. Compared
to their analysis, we update several parameters like the Solar distance to
the Galactic center, which requires
a re-fitting of the model parameters. More importantly, we add new elements
like the Local Arm and a devoid region of the Sagittarius-Carina Arm, as well
as using information on observed nearby \OBA s.
The resulting model can be used for reproducing the N\,II intensity in the
Galactic plane, and for simulating OB associations and past SNe in the
Milky Way. In addition to the use of H\,II regions as source distribution
for Galactic CRs, pulsars or black holes, accounting
for H\,II regions might be useful in the interpretation of radio dispersion
and scattering measurements as electron density tracers~\cite{Ocker:2024hpl}.
The complete model is made publicly available in the python code {\tt SNOB},
available at \url{https://github.com/VimiZap/SNOB}.

\section{Modelling the N\,II intensity of the Galactic plane}

Our aim in this section is to reproduce the
N\,II 205$\mu$m intensity measured in the Galactic plane by the FIRAS
instrument on-board the COBE satellite~\cite{Fixsen_1999}. After a brief
discussion of the data and our fitting procedure, we first introduce an
axial-symmetric three-dimensional model which will then serve as the
foundation for a more realistic source model based on spiral arms.

\subsection{FIRAS data and fitting procedure} \label{FIRAS}

We want to use the longitudinal profile of the N\,II intensity in the Galactic
plane as shown in Fig.~5(e)
of Ref.~\cite{Fixsen_1999} to calibrate our source model for the Galactic
N\,II emission. For a given spatial distribution of the isotropic
N\,II line emissivity $\eps(\vec x)$ in the Milky Way, the intensity $I$
in the direction specified by the Galactic longitude $l$  and latitude $b$
can be calculated from the line-of-sight integral
\begin{equation}
 I(l, b) = \frac{1}{4\pi} \int_0^\infty \d r \, \eps(r,l,b) .
    \label{eq: axisymmetric non-averaged intensity}
\end{equation}
Here, $r$ denotes the distance from the Sun, while $r_{\max}$ is the maximal
distance of Galactic N\,II sources. Moreover, we use that absorption can be
neglected for this line.

The FIRAS beam approximately was a top hat with an opening angle of $7^\circ$,
and thus the instrument was not
able to resolve the Galactic plane.  As a simple alternative to the
procedure used in Ref.~\cite{Fixsen_1999} to derive the intensity $I(l,b=0)$
in the Galactic plane, we use the following recipe suggested by Fixsen~\cite{F}:
We average our modelled intensities over \(\Delta l = 5^\circ\) in longitude,
while we integrate them over \(b=\pm 5^\circ\) in latitude.
Then we 
divide by \(\Delta b = 1^\circ\), assuming that all radiation from the
Galactic plane is contained within $|b|<0.5^\circ$.
Thus we compare the average
intensities $\left< I(l, 0) \right>$ of our models in the Galactic plane
defined by
\begin{equation}
  \left< I(l, 0) \right> = \frac{1}{\Delta l \, \Delta b}
  \int_{b = -5^\circ}^{b = +5^\circ} \d b  \cos b \int_{l = -2.5^\circ}^{l = +2.5^\circ} \d l
  \int_{0}^{\infty} \d r \,
  \frac{\eps(r, l, b)}{4 \pi} 
\end{equation}
to the longitudinal profile $I(l,0)$ of the FIRAS data shown in
Ref.~~\cite{Fixsen_1999}.
In order to fit the parameters of the source models, we employ a \(\chi^2\)
test,
\begin{equation}
    \chi^2 = \frac{1}{N-m} \sum_{i=1}^{N} \frac{(O_i - C_i)^2}{\sigma_i^2}.
\end{equation}
Here, \(C_i\) are the modelled and \(O_i\) the observed intensities,
respectively, while 
\(\sigma_i\) is the error of the observed intensity $C_i$. In the final
version of the model, the number of fit parameters amounts to \(m = 40\)
while the number of data points is \(N = 1800\). This large number of
parameters makes the testing of
all different combinations unpractical. Instead, the parameters used in
Ref.~\cite{higdon_galactic_2013} were taken as a starting point and
adjusted first by eye until the model became reasonable as compared to the
FIRAS data. Then subsets of parameters were fitted within a chosen
range around their initial value.

\begin{figure}[h!]
    \centering
    \includegraphics[width=0.8\linewidth]{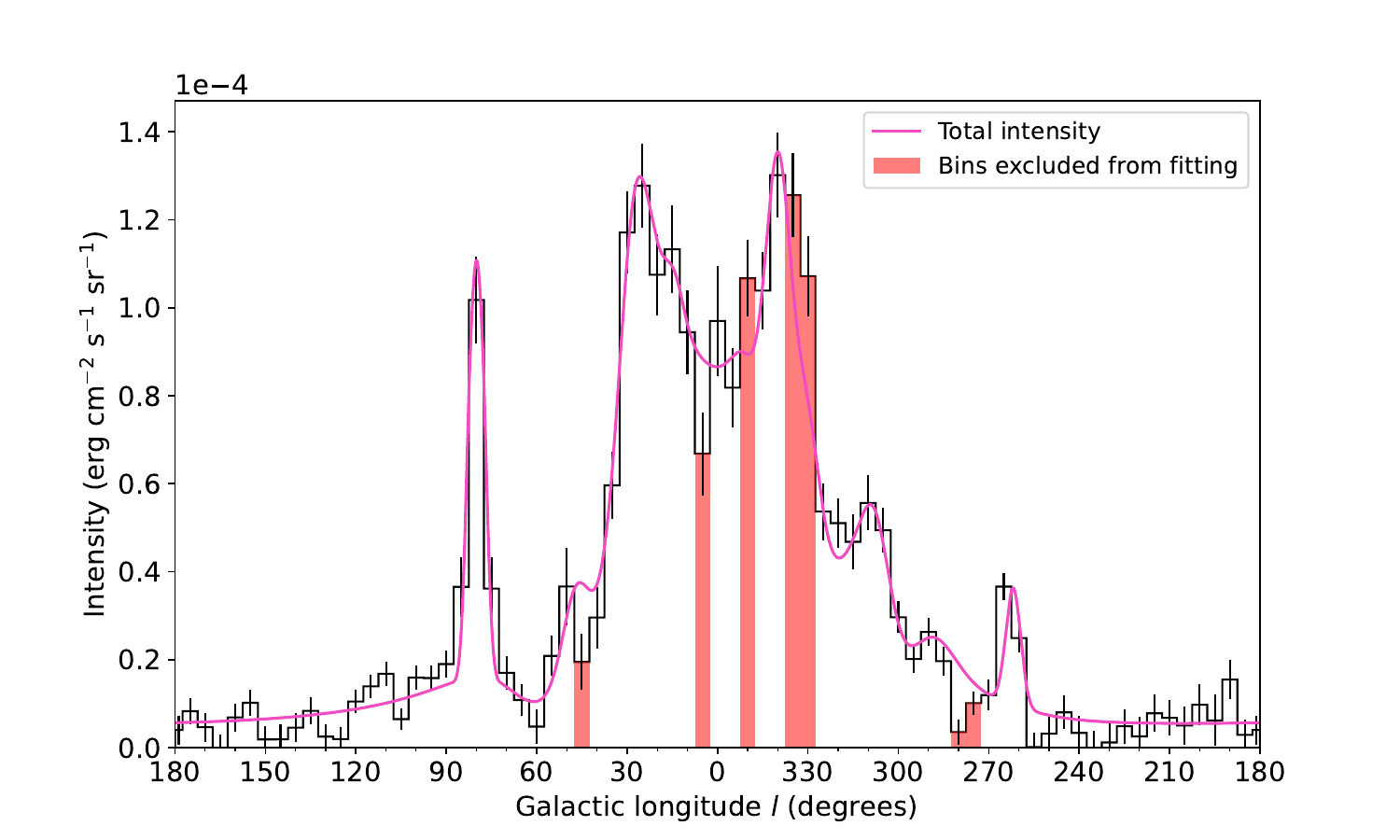}
    \caption{N\,II $205\mu$m line intensity (magenta line) in the Galactic
      plane for the best-fit model as  function of the Galactic longitude
      together with the measured N\,II line intensity
      (black histogram) from FIRAS in Figure 5(e) of \cite{Fixsen_1999};
      the bins removed from the fitting procedure are highlighted in red.
    \label{fig: bins_excluded_from_fitting}
    }
\end{figure}

Even though the final model has \(40\) parameters, it remains relatively
simple compared to the complexity of the Milky Way, and thus there are several
regions where the model severely over- or underestimates the N\,II intensities.
As the deviations of the model from the observed data are so prominent
at specific bins, the algorithm for optimising the parameters would
favour parameters which minimise these large deviations at the
expense of other features of the spiral arms. Examples of this behaviour
are the skewing of the Sagittarius-Carina Arm to
minimise the gap at \(l = {280}^{\circ}\) at the expense of its spiral arm
tangents, and favouring a significant fractional contribution for
Norma-Cygnus at the expense of Perseus to minimise the gap at
\(l = {330}^{\circ}\) and \(l = {350}^{\circ}\). From observations, it is
known that Perseus and Scutum-Crux are the most luminous arms, rendering
the latter solution incompatible with observations. 
Therefore, several bins were removed from the calculation of the
reduced \(\chi^2\) during the  optimisation of the parameters to keep
the model reasonable compared to observations. These were the bins at
$l = \{ {5}^{\circ}, {45}^{\circ}, {275}^{\circ}, {280}^{\circ}, {330}^{\circ},
{335}^{\circ},350^{\circ}\}$ which are highlighted in
Fig.~\ref{fig: bins_excluded_from_fitting}.
However, all the reduced \(\chi^2\) values given latter 
are calculated from all the data points---thus the listed bins were only left
out for the fitting process.

\subsection{Axial-symmetric model} \label{ch: axisymmetric model}

The simplest approach to model the Galactic N\,II emissivity is to approximate
the distribution of sources in the Milky Way as a cylinder and to assume
axial symmetry. To convert from Galactic to Galactocentric cylindrical
coordinates, we use
\begin{subequations}
  \begin{align}
    \rho &= \sqrt{x^2 + y^2} = \sqrt{r^2 \cos^2 b + R_\odot^2 - 2 R_\odot r \cos b \cos l}, \\
    \theta &=\arctan(y/x) = \arctan\left(\frac{R_\odot - r \cos b \cos l}{r \cos b \sin l} \right), \\
    z &= r \sin b,
  \end{align}
  \label{eq: earth_centric to galacto_centric}
\end{subequations}
with $R_\odot= 8.2$\,kpc as the Solar distance to the Galactic center, as
suggested by the results of the GRAVITY
Collaboration~\cite{abuter_geometric_2019}.
In addition to axial symmetry, we assume that the radial and the
vertical dependence of the
N\,II 205$\mu$m line emissivity factorise, such that we can set
\begin{equation}
    \eps(\rho,z) = L S(\rho) P(z) / A.
    \label{eq: axisymmetric emissivity}
\end{equation}
This excludes effects as a warp in the Galactic disk, which is most
prominent in the outer Galaxy. In Eq.~(\ref{eq: axisymmetric emissivity}),
\(L\) is the total Galactic N\,II $205\mu$m luminosity,  $S(\rho)$
is the (dimensionless) radial and \(P(z)\) the (dimensionfull) vertical
dependence of the N\,II source distribution, while \(A\) denotes the
effective Galactic disk area.
%
%
For the vertical distribution  $P(z)$ of N\,II sources we employ
following Ref.~\cite{higdon_galactic_2013}  a Gaussian distribution,
\begin{equation}
    P(z) = \frac{\e^{-z^{2}/2\sigma_z^{2}}}{\sqrt{2\pi}\sigma_z} .
    \label{eq: height_distribution}
\end{equation}
Because of the large beam size of the FIRAS instrument, the FIRAS data
cannot be used to determine the width  \(\sigma_z\) of this
distribution. Instead, one can use hydrogen radio recombination lines
(RRL) as a tracer for the distribution of H\,II regions, which
are known to stem from their outer
envelopes~\cite{Anantharamaiah_1986, Roshi_2001}.
Reference~\cite{Roshi_2001} showed that the low-frequency RRL have a
full width at half maximum (FWHM) of about \(1.8^\circ\) in latitude.
They also found
that \(75 \%\) of the identified H\,II regions in high-frequency RRL surveys
were located in the Galactic plane, with $|b| < 1.5^\circ$.
Moreover, a more recent study~\cite{RRL_2006} of RRLs in 106 Galactic H\,II
regions found
only four regions with  a Galactic latitude greater than $4^\circ$, while
88 were within \(|b| < 1^\circ\), equating to about \({74}{\%}\) of the
sample size. These studies substantiate the argument that H\,II regions are
strongly concentrated in the Galactic disk. 
Reference~\cite{higdon_galactic_2013} estimated that  $\sigma_z = 0.15$\,kpc
reproduces the FWHM reported by \cite{Roshi_2001}.
Even though this value is much greater than the exponential scale height
of OB stars of $0.045$\,pc \cite{Reed_2000} and the scale height of molecular
clouds of $0.035$\,kpc \cite{stark_lee_2005}, it is comparable to the
estimate in Ref.~\cite{MWL1997}, where $\sigma_z = 0.15$\,kpc as value
for the vertical scale height \(\sigma_z\) of  H\,II regions is reported.
This is the value we adopt also in this work for N\,II sources.


For the (dimensionless) radial distribution  $S(\rho)$ of N\,II sources we
employ following Ref.~\cite{MWL1997} an exponential distribution,
\begin{equation}
  S(\rho) \propto \text{e}^{-\rho/H^{\text{N\,II}}_{\rho}} ,
  \label{eq: axisymmetric_distribution}
\end{equation}
for $\rho_{\min}\leq\rho\leq\rho_{\max}$. We use
$\rho_{\text{min}} = 3$\,kpc  and $\rho_{\text{max}} = 11$\,kpc, corresponding
to the boundary of the  H\,II regions listed in 
Ref.~\cite{Smith_biermann_mezger_1978}.
The region $\rho<\rho_{\text{min}} = 3$\,kpc corresponds roughly to the
extension of the Galactic bar, where star formation is suppressed.
In order to determine the scale length $H^{\text{N\,II}}_{\rho}$,
we have to account for the radial N/H gradient which biases the radial N\,II
and H\,II distributions. Assuming the same ionisation degree and for all
three distributions an exponential
dependence, $\propto \exp(-\rho/H_i)$ with scale length
$H_i=\{H^\text{H\,II},H^\text{N\,II},H^\text{N/H}\}$, they
are connected by
\begin{equation}
  \frac{1}{H^{\text{N\,II}}_{\rho}} \simeq
  \frac{1}{H^{\text{N/H}}_{\rho}} + \frac{1}{H^{\text{H\,II}}_{\rho}}. 
    \label{eq: NII scale length relation}
\end{equation}
Reference~\cite{Esteban_garcia_2018} recently revisited the radial abundance
gradients of nitrogen and oxygen. They selected \(13\) H\,II regions located at
Galactocentric distances between 5.7\,kpc and 16.1\,kpc. Because of their
low degree of ionisation, the authors of Ref.~\cite{Esteban_garcia_2018}
argued that their number for the nitrogen radial abundance gradient is the
first which is independent of the ionisation correction factor. Their
reported change is
$\log({\rm N/H})=(-0.059 \pm 0.009)\rho/{\rm kpc}+{\rm const}$.
Inserting \(10^{0.059} \simeq 1.15\) per kpc
together with $H^{\text{H\,II}}_{\rho} =3.5$\,kpc from Ref.~\cite{MWL1997}
into Eq.~(\ref{eq: NII scale length relation}),
we obtain $H^{\text{N\,II}}_{\rho} = 2.4$\,kpc.
Finally, the effective Galactic disk area $A$ is defined as the disk area
weighted by the disk emissivity, 
\be
A =\int\d\theta\int\d\rho\rho \:S(\rho)=
 2\pi \left(H^{\text{N\,II}}_{\rho}\right)^2
        [(1 + \rho_{\text{min}}/ H_\rho) S(\rho_{\text{min}}) - (1 + \rho_{\text{max}}/ H_\rho ) S(\rho_{\text{max}}))],
\ee
which with the adopted values for \(\rho_{\text{min}}\), \(\rho_{\text{max}}\)
and \(H^{\text{N\,II}}_\rho\) evaluates to $A=21.265$\,kpc$^2$.
%

\begin{figure}[h!]
    \centering
    \includegraphics[width=0.80\linewidth]{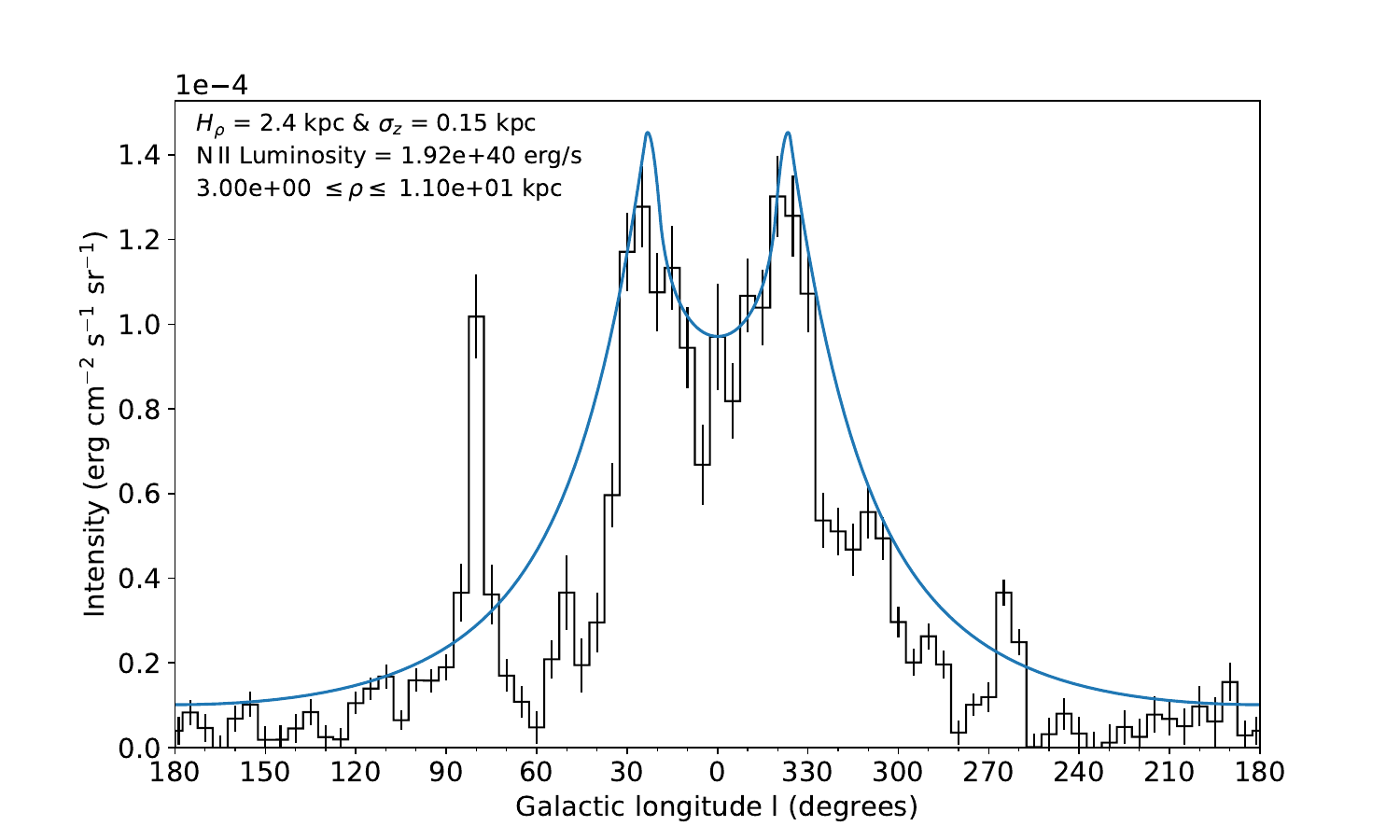}
    \caption{N\,II $205\mu$m line intensity (blue line) as function of the Galactic longitude in the axial-symmetric model together with the measured N\,II line intensity
       (black histogram) from FIRAS in Figure 5(e) of \cite{Fixsen_1999}.
    }
    \label{fig: axisymmetric_nii_intensity}
\end{figure}

Following Ref.~\cite{higdon_galactic_2013}, this axial-symmetric model is then
normalised to the observed FIRAS N\,II  intensity at the longitude
\(l = 30^{\circ}\).  Our best-fit model shown in
Fig.~\ref{fig: axisymmetric_nii_intensity} together with the observed
intensities from Ref.~\cite{Fixsen_1999}  requires a total
Galactic N\,II
luminosity of  $1.9\times 10^{40}$erg/s, which is in close agreement with the
value $1.8\times 10^{40}$erg/s obtained by Higdon and Lingenfelter.
In order to check how our results depend on the
value of $\rho_{\text{min}}$, we increased $\rho_{\text{min}}$ from 3 to 5\,kpc.
As a result, the peak in the intensity moves to $l\simeq 45^\circ$ for
$\rho_{\text{min}}=5$\,kpc, in contradiction to observations.
We conclude therefore that a value close to our default value
$\rho_{\text{min}}=3$\,kpc is favoured by the N\,II data.
From the figure, it is however clear that the axial-symmetric model poorly
reproduces
the observed intensities outside the Galactic center region, as it
falls off too slowly and fails to capture many of the structures visible
in the data.
Hence, a more complicated and accurate model is needed and will be constructed
next. 

\subsection{Spiral arm model} \label{ch: spiral arm model}

We consider next a model with four spiral arms, adding then the Cygnus
region and the Gum nebula which are clearly visible local H\,II regions.
Moreover, we introduce the Local Arm and an underdense region in the
Sagittarius-Carina Arm.

\subsubsection{Four-arm model}

\begin{figure}[h!]
  \centering
  \includegraphics[width=0.42\linewidth]{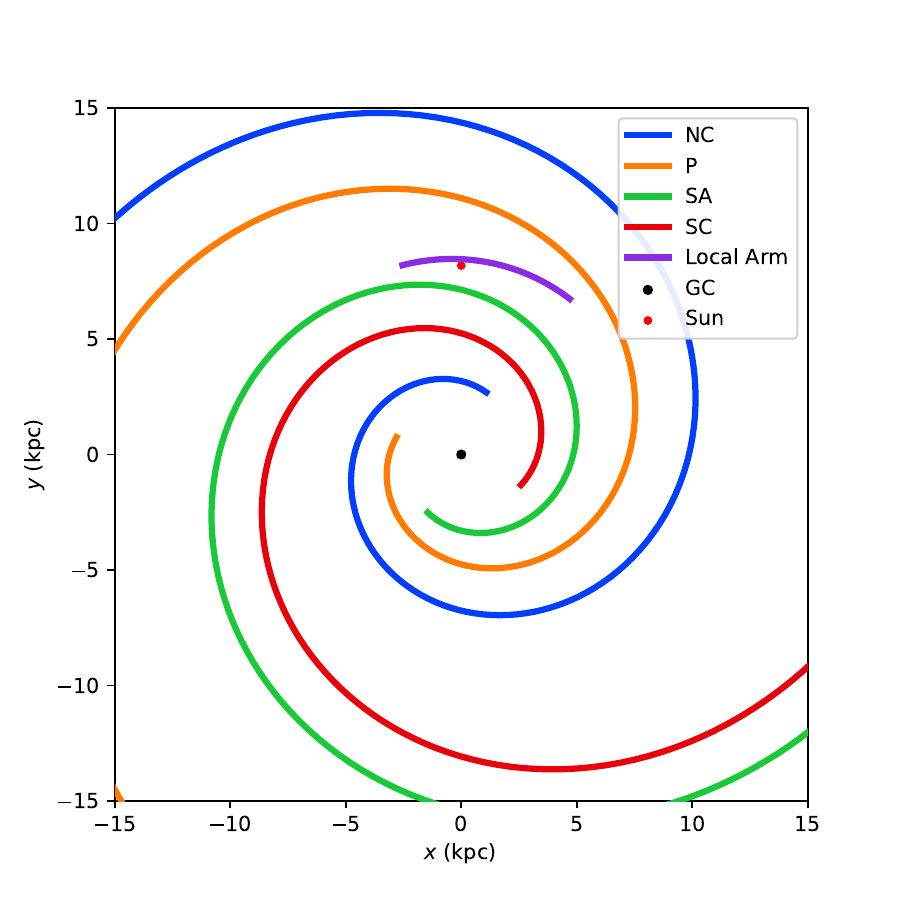}
  \includegraphics[width=0.57\linewidth]{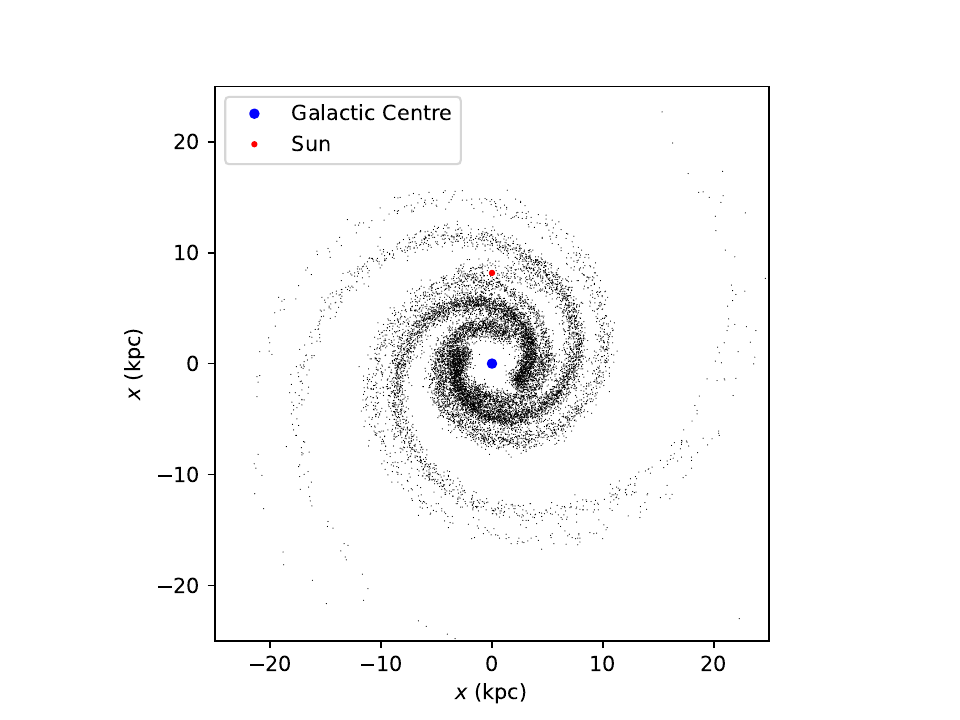}
  \caption{Left: Schematic view of the Milky Way, showing the four major spiral arms Norma-Cygnus (NC), Perseus (P), Sagittarius-Carina (SA), and Scutum-Crux (SC), as well as the Local Arm. The Galactic center (GC) and the Sun are also shown. Right: Modelled Galaxy after 100\,Myrs. Each black dot represents an OB association with at least \(2\) OB stars.
        \label{fig: schematic spiral arm structure} }
\end{figure}

Studying H\,II regions and their massive ionising stars,
Ref.~\cite{Georgelin_1976}
proposed that the Galaxy consists of four spiral arms known as Norma-Cygnus
(NC), Perseus (P), Sagittarius-Carina (SA), and Scutum-Crux (SC), for
more recent works on the spiral arm structure of the
Milky Way see Refs.~\cite{vallee_2008, vallee_2013,hou_han_2014}. 
Both Refs.~\cite{vallee_2008} and \cite{hou_han_2014} use a logarithmic model
for the positions $\rho (\theta)$ of the spiral arm medians in the
Galactocentric cylindrical coordinate system, where
$\rho (\theta) = \rho_{\text{min}} \e^{k(\theta - \theta_{0})}$ for
$\rho_{\text{min}}\leq \rho \leq \rho_{\text{max}}$ with
$k = \tan(p)$.
Here, \(p\) is the pitch angle for the given spiral arm,
\(\rho_{\text{min}}\) and \(\rho_{\text{max}}\) are respectively the inner and
outer limits of the spiral arm, and \(\theta_{0}\) is the starting angle for
the spiral arm as measured counter-clockwise from the \(x\) axis.
Following Refs.~\cite{vallee_2008,Cameron_2010,higdon_galactic_2013},
we will use 
$\rho_{\min} =2.9$\,kpc and $\rho_{\max} = 35$\,kpc for the inner and
outer limit of the spirals. Since we use the more up-to-date value of
$R_\odot = 8.2$\,kpc, the starting angles and pitch angles have been refitted
from the FIRAS data. The obtained values are listed in
Tab.~\ref{table12} in units of degrees. 
The resulting four-arm spiral model is shown in the left panel of
Fig.~\ref{fig: schematic spiral arm structure} together with the Local Arm,
which will be discussed later. In the right panel, we 
show the distribution of \OBA s, which follows the same spiral arm structure
as we will discuss in section~3.

Following Refs.~\cite{Taylor_1993, Cordes_2002}, the
perpendicular extension of the arms is
modelled by a Gaussian,
\begin{equation}
    P_{\Delta}(\Delta) = \e^{-\Delta^{2} / 2\sigma_{\text{A}}^{2}}.
    \label{eq: transverse density spiral arms}
\end{equation}
Note that the transverse densities $P_{\Delta}$  are normalised to one at
$\Delta=0$ such that they give the relative density
with respect to the arm median density, which is given by 
\begin{equation}
    P_{\rho}(\rho) = \e^{-\rho/H_{\rho}}.
    \label{eq: arm median density}
\end{equation}
In addition, Eq.~(\ref{eq: transverse density spiral arms}) is used to
determine the fall-off in density at the end of the spiral arms,
where the calculated densities are projected on a circular arc of $180^\circ$
around the final point. This ensures
a smooth transition at the endpoint to the surrounding medium.
As in the axial-symmetric model of section~\ref{ch: axisymmetric model}, the
distribution vertical to the Galactic plane is given by
Eq.~(\ref{eq: height_distribution}) with
$\sigma_z = 0.15$\,kpc. The relative planar density of each spiral arm is given
by the product \(P_{\rho}(\rho) \, P_{\Delta}(\Delta)\).
The source-weighted area of the spiral arm $i$ becomes
\begin{equation}
  A_i = \int_{0}^{2\pi}\d \theta \int_{0}^{\infty} \d\rho\rho \,
      P_{\rho}(\rho_j(\rho, \theta)) P_{\Delta}(\Delta) ,
\end{equation}
where \(\rho_j\) is a given point along the spiral arm median as a function
of \(\rho\) and \(\theta\). The function
\(P_{\Delta}\) remains the same for each spiral arm and every \(\rho_j\), as
the transverse density always is relative to the arm median density.

\begin{table}[tb]
    \centering
    \begin{tabular}{@{}lrrrrrrrrrrrr@{}}
     & \(\theta_{0, \text{NC}}\) & \(\theta_{0, \text{P}}\) & \(\theta_{0,\text{SA}}\) & \(\theta_{0, \text{SC}}\) & \(p_{\text{NC}}\) & \(p_{\text{P}}\) & \(p_{\text{SA}}\) & \(p_{\text{SC}}\)  & \(f_{\text{NC}}\) & \(f_{\text{P}}\) & \(f_{\text{SA}}\) & \(f_{\text{SC}}\)\\
    our fit & 68 & 165 & 240 & 333 & 13.5 & 15.1 & 13.8 & 16.2 & 0.19 & 0.35 & 0.16 & 0.29\\
    Ref.~\cite{higdon_galactic_2013} & 70 & 160 & 250 & 340 & 13.5 & 13.5 & 13.5 & 15.5 & 0.18 & 0.36 & 0.18 & 0.28\\
    \end{tabular}
    \caption{The starting and pitch angles in degrees, plus the relative
      weight of
      the four spiral arms from our fit to the FIRAS data compared to the values of Ref.~\cite{higdon_galactic_2013}.
      \label{table12}}
\end{table}

Finally, each of  the four spiral arms is given a relative weight \(f_i\) to
reproduce the observational data more accurately. This is in line with, e.g.,
Ref.~\cite{Drimmel_Spergel_2001} who fitted a three-dimensional model of
the far-infrared (FIR) and near-infrared (NIR) data measured by COBE/DIRBE and
found that the spiral arms Scutum-Crux and Perseus dominated the NIR emission.
To improve their model, they had to reduce the Sagittarius-Carina Arm's
luminosity
relative to the other arms. Other works such as Refs.~\cite{Benjamin_2005, Benjamin_2008, Hou_han_2015} also support that the Sagittarius-Carina arm is weaker
than the other arms. Reference~\cite{Cameron_2010} also finds that the Scutum-Crux
and Perseus arms have significantly stronger emissions than the Sagittarius-Carina
and Norma-Cygnus arms. 
As for the spiral arm angles and pitch angles, the values given by
\cite{higdon_galactic_2013} have to be tweaked to fit with the new parameters for
the Galaxy, and the obtained values can be found in Table~\ref{table12}.
Note that the sum of all fractional contributions only adds to \(0.99\).
This is because we anticipate the inclusion of the Local Arm, which will be
described in subsection~\ref{ch: local arm}.

\subsubsection{Cygnus region}

The Cygnus region is a massive star forming region of our Galaxy, containing
several OB associations and hundreds of OB
stars~\cite{Quintana_Wright_2021, Wright2020}. The most massive of these
associations is the well-known Cyg~OB2 association, which contains at least
\(50\) O stars. Cyg~OB2 is embedded in the large star forming region called
Cygnus X located at \( l\simeq 80^\circ\), \(b\simeq {0.8}^{\circ}\) \cite{Wright_2015, Wright2020, Rygl_2012}. with an angular extension of
\(\simeq 10^\circ\)~\cite{Reipurth_schneider_2008} and a distance of
$\simeq 400$\,pc.
The bright feature in the FIRAS data at \(l\simeq 80^\circ\) will thus be
interpreted as the star forming region Cygnus X. Following
Ref.~\cite{higdon_galactic_2013}, this feature is modelled as a uniform
sphere located at $(l,b) = (80^\circ, 0^\circ)$ with a radius of $R=75$\,pc,
corresponding to an angular width of $\simeq 12^\circ$.
To mimic the effect of the \({7}^{\circ}\) FIRAS beam, the modelled intensities
from Cygnus X were weighted with a Gaussian distribution with a full width at
half maximum of \({7}^{\circ}\). In the same manner as for the spiral arm
intensities, the contribution from the Cygnus X region was averaged over
\(\Delta l = {5}^{\circ}\) and \(\Delta b \simeq {5.9}^{\circ}\), with
\(\Delta b\) being the angular extent of the source. The fitted N\,II
luminosity for Cygnus~X is $2\times 10^{37}$erg/s, 
identical to the value obtained in Ref.~\cite{higdon_galactic_2013}.

\subsubsection{Gum Nebula}

\begin{figure}[h]
    \centering
    \includegraphics[width=1\linewidth]{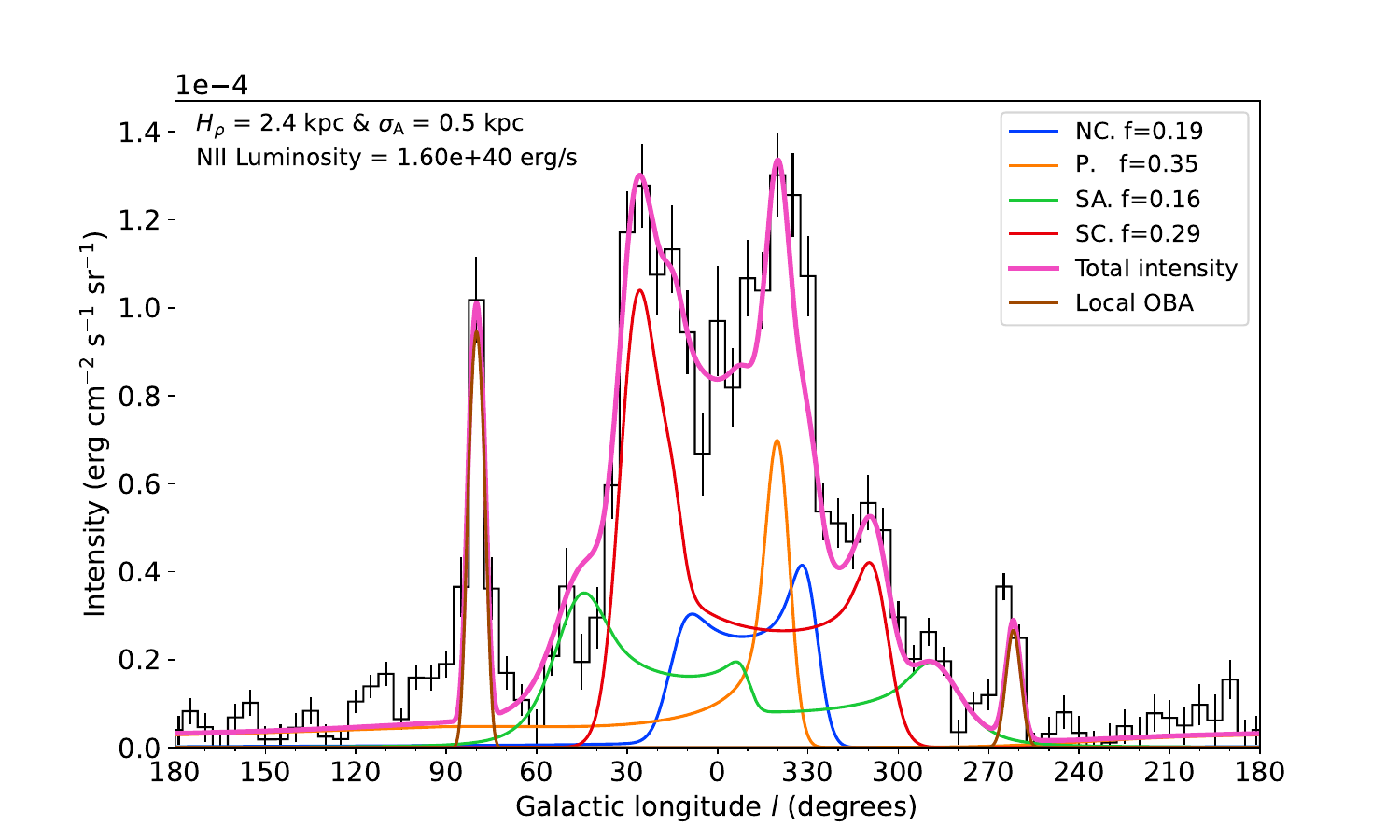}
    \caption{N\,II line intensity for the four spiral arm model plus the
      Cygnus~X and the Gum nebula.
      The measured N\,II line from FIRAS in Figure 5(e) of \cite{Fixsen_1999} is shown as a histogram. 
    }
    \label{fig: spiral_arm_nii_intensity_four_arms_gum_cygnus}
\end{figure}

The bright feature in the FIRAS data at \(l \simeq {260}^{\circ}\) is
interpreted as the H\,II region known as the Gum Nebula, first reported in
Ref.~\cite{Gum_1952}. This H\,II region is primarily excited by the massive
runaway O4I star \(\zeta\) Puppis and the binary \(\gamma^2\) Velorum in
the Vela OB2 association, the latter being composed of a massive O star and
a Wolf-Rayet star \cite{Reynolds_1976}.
The most massive and brightest member of Vela~OB2 is \(\gamma^2\) Velorum
which is located at \(l = {262.8}^{\circ}\), \(b = {-7.7}^{\circ}\) \cite{o_star_catalogue_2004} at a distance of $368^{+38}_{-13}$\,pc~\cite{Millour_2007}.
Following Ref.~\cite{higdon_galactic_2013}, the feature in the FIRAS data at \(l \simeq {260}^{\circ}\) is modelled as a single spherical source located
at \(l={262}^{\circ}\), \(b = {0}^{\circ}\) and a distance of 330\,pc. As for Cygnus X, the Gum Nebula will be weighted over a normal distribution with a full width at half maximum of \({7}^{\circ}\) to simulate the effect of the FIRAS beam. It will also be averaged over \(\Delta l = {5}^{\circ}\) and \(\Delta b \simeq {10.4}^{\circ}\).
The required N\,II luminosity of the Gum Nebula is $1.2\times 10^{36}$erg/s,
which is consistent with the value  obtained in
Ref.~\cite{higdon_galactic_2013}.
The effect of including Cygnus~X  and the Gum Nebula to the N\,II intensity
profile can be seen in
Fig.~\ref{fig: spiral_arm_nii_intensity_four_arms_gum_cygnus}.

\subsubsection{Local Arm} \label{ch: local arm}

In addition to the four major arms,
star forming regions between the Perseus and Sagittarius-Carina Arms have
been known for a long time. Reference~\cite{hou_han_2014} recently investigated
the
spiral structure of the Milky Way, employing data from more than 2500 known
H\,II regions, 1300 giant molecular clouds (GMCs), and 900 6.7 GHz methanol
masers. They found that a model with four major spiral arms and a Local Arm
best fitted the data.  The Local Arm begins near the Perseus  Arm and
extends past the Sun towards the Sagittarius-Carina  Arm. The parameters
from Ref.~\cite{hou_han_2014} for the Local Arm together with a value for the
fractional contribution and the result from the reduced \(\chi^2\) fitting
are given in Table~\ref{table3}. We use the values
for \(\theta_{0, \, \text{LA}}\) and \(p_{\text{LA}}\) given in
Ref.~\cite{hou_han_2014}, while $\theta_{\max,\rm LA}$ and
$f_{\rm LA}$ were fitted.

\begin{figure}[h!]
    \centering
    \includegraphics[width=1\linewidth]{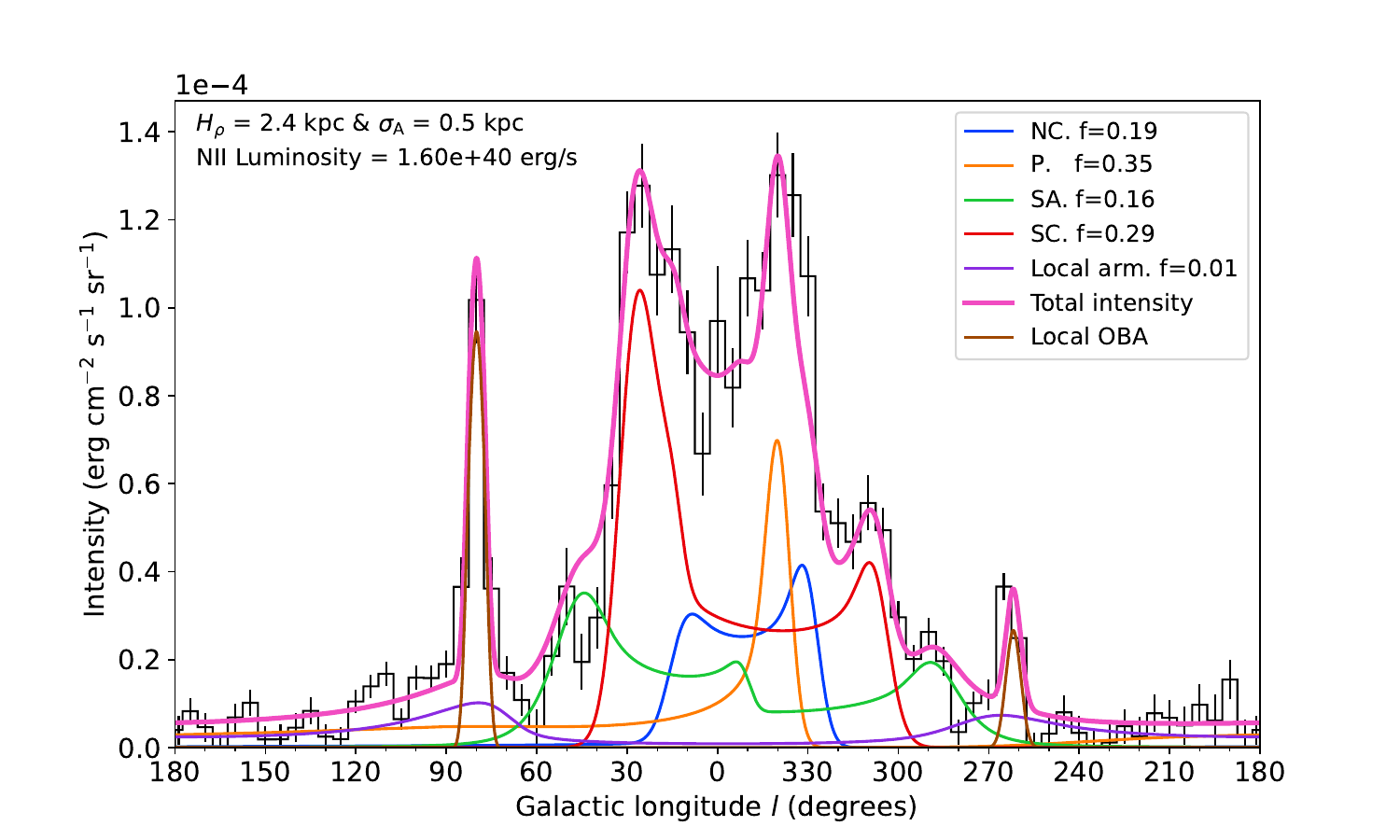}
    \caption{N\,II line intensity for the four spiral arm model adding two local sources and the Local Arm. 
      The measured N\,II line from FIRAS in Figure 5(e) of \cite{Fixsen_1999} is shown as a histogram.
    }
    \label{fig: spiral_arm_nii_intensity_five_arms_gum_cygnus}
\end{figure}

\begin{table}[tb]
    \centering
    \begin{tabular}{@{}lrrrr@{}}
    Name & \(\theta_{0, \, \text{LA}}\) & \(\theta_{\text{max}, \, \text{LA}}\) & \(p_{\text{LA}}\) & \(f_{\text{LA}}\) \\
    Hou \& Han & 55.1 & 110 & 2.77 &  \\
    our fit & 55.1 & 107 & 2.77 & 0.01 \\
    \end{tabular}
    \caption{Starting and maximal angle, and the relative weight of the Local
      Arm,
      from our fit compared to those of Ref.~\cite{hou_han_2014}.
      \label{table3}}
\end{table}

The result of adding the Local Arm to the model can be seen in
Fig.~\ref{fig: spiral_arm_nii_intensity_five_arms_gum_cygnus}. The most
notable contributions to the resulting N\,II line intensity are the regions
\(l \simeq {70}^{\circ}\) and \(l \simeq {110}^{\circ}\). The inclusion of the
Local Arm also contributes towards the Gum Nebula, increasing the total
modelled N\,II intensities to match the observed data. The inclusion of the
Local Arm lowered the reduced \(\chi^2\) from \(16.72\) to \(12.47\). Contrary
to Ref.~\cite{higdon_galactic_2013}, we thus conclude that including the
Local Arm improves the model considerably.

\subsubsection{Devoid region of the Sagittarius-Carina Arm}

\begin{figure}[h!]
    \centering
    \includegraphics[width=0.95\linewidth]{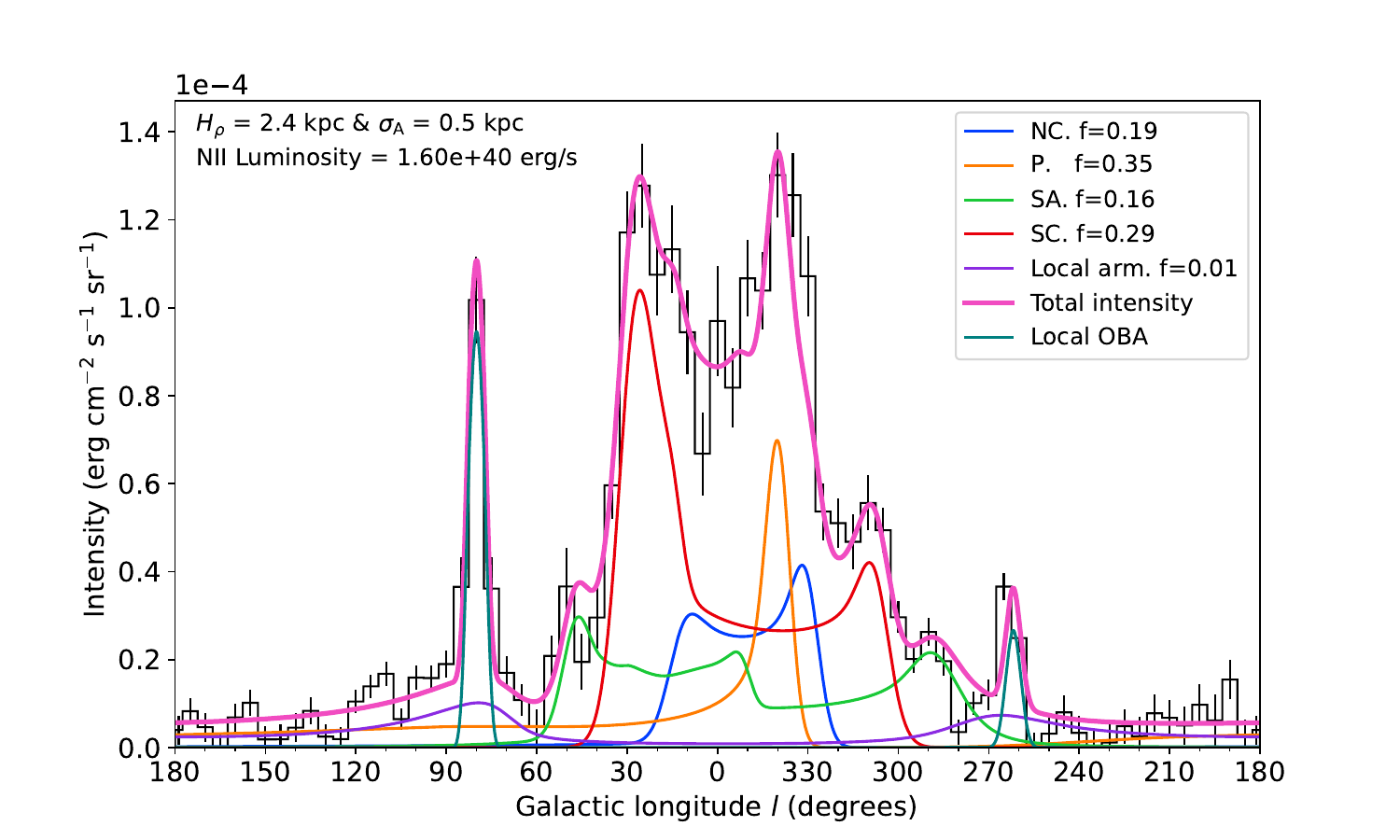}
    \caption{N\,II intensity
      for the four spiral arm model adding two local sources, the Local Arm and
      the devoid region of Sagittarius.
       The measured N\,II line from FIRAS in Figure 5(e) of \cite{Fixsen_1999} is shown as a histogram.
    \label{fig: modelled_intensity_all_contributions}
    }
\end{figure}

Looking at Fig.~\ref{fig: spiral_arm_nii_intensity_five_arms_gum_cygnus}, it
is clear that the model is overestimating the
intensities in the range \(l = {35}^{\circ}\) to \(l = {65}^{\circ}\).
This portion of the Sagittarius-Carina Arm has been known for a long time
to be devoid of optical tracers and to suffer from heavy
extinction~\cite{Forbes_1983, Forbes_1984, Forbes_1985}.
We select a specific portion of the arm in this direction to model this devoid
region, reducing the corresponding \(\sigma_{\text{d}}\). This effectively
leads to a faster  fall-off of the density surrounding to the arm median,
instead of simply decreasing the modelled densities of the entire arm segment. 
As shown in Fig.~\ref{fig: modelled_intensity_all_contributions}, the
resulting Sagittarius-Carina Arm matches the observed intensities in this
region better. The devoid region is defined by the parameters
$\rho_{\text{min}}=5.1$\,kpc,
$\rho_{\text{max}}=7.0$\,kpc and $\sigma_{\text{d}}=0.25$\,kpc, where
\(\rho_{\text{min}}\) determines the closest distance from the Galactic center
to the devoid region and \(\rho_{\text{max}}\)  the other edge
further away from the Galactic centre. After incorporating the devoid region,
the overall fit for the model improves, lowering the reduced \(\chi^2\)
from \(12.47\) to \(11.25\).

\subsubsection{Summary}

Our final model for the N\,II emissivity consists of the seven components
described above: Four logarithmic spiral arms with individual weights,
where we include in the case of the Sagittarius-Carina Arm an underdense
region, plus the Local Arm. In addition, the Cygnus region and the Gum nebula
as the two most important local H\,II regions were taken into account.
The  N\,II intensity of these individual contributions and the resulting
total intensity is shown in
Fig.~\ref{fig: modelled_intensity_all_contributions}.

\section{Supernova progenitor distribution}

The model for the Galactic N\,II emission developed in the previous
section will be the basis for the spatial probability distribution of 
\OBA s and SNe in the Milky Way.
In particular, we will reinterpret expressions like
$P_\Delta(\Delta)$ or $P_\rho(\rho)$ as probability distributions for the
spatial density of \OBA s.
In order to make the model in our local neighbourhood more accurate,
we add  those  observed \OBA s where we have reliable information on
their distance, mass and age.

\subsection{Connecting  OB associations and supernovae}
\label{OBSB}

\subsubsection{SNe per associations}

The number of newly born associations and their stellar content should reflect
the expected SNe birthrate, which is about three per
century~\cite{Rammann_1994, van_den_Bergh_1994}, out of which about
80\%--90\% are core-collapse SNe of type II and Ib/c. 
How the mass of the \OBA\ is  distributed among different stars is
determined by the stellar initial mass function (IMF).
We will here use the modified Kroupa IMF, where the index \(\alpha_3\) has
been increased from 2.35 to 2.7 to take into account unresolved
binaries~\cite{Kroupa2002}.
Combining then the size-frequency distribution
of young stellar clusters, the modified Kroupa IMF and the study of ionising
luminosities in Ref.~\cite{MWL1997}, the number of
clusters born per Myr with a number of stars \(N_*\) between \(N_*\) and
\(N_* + \d N_*\) was derived in Ref.~\cite{Higdon_lingenfelter_2005} as
\begin{equation}
    \frac{\d n_{\text{sc}}(N_*)}{\d N_* \, \d t} = 1.65 \left (\frac{{1.7\times 10^{6}}{}}{N_*^{1+\alpha}} \right) {}{/{\rm Myr}}, \ \text{for} \ 190 \leq N_* \leq {1.7\times 10^{6}}{}.
    \label{eq: star cluster star distribution}
\end{equation}
Here, \(n_{\text{sc}}(N_*)\) is the number of clusters with \(N_*\) stars of all
masses. The slope \(\alpha\)  of this power law has to be determined from
observations of H\,II regions: The authors of Ref.~\cite{Kennicutt_1989}
investigated 30 nearby galaxies and found as best fit value
\(\alpha = 1 \pm 0.5\),
while Ref.~\cite{MWL1997} obtained \(\alpha = 0.99 \pm 0.25\)
focusing on \OBA s in the Milky Way and  excluding the smallest H\,II
regions due to incompleteness in the dataset on the lower end.
A value close to one is also theoretically expected, because otherwise
massive stars would be concentrated either in the smallest or largest
associations. We will use as our default parameter \(\alpha = 0.8\),
a value within the reported uncertainties which fits better to
the observational data on OB associations than \(\alpha = 1\),
as we will see later.

On average, the number of SN progenitors (SNPs)  born is related to the total
number of
stars formed. Denoting by \(f_{\text{SN}}\) the fraction of formed stars 
with initial mass\footnote{Note that mass transfer in binary
    systems can change this simple picture.}
above \(8 \, M_{\odot}\), which is the commonly
stated minimum mass for core-collapse
SNe~\cite{Woosley__weaver_1995, Smartt_2009},
the modified Kroupa IMF gives  \(f_{\text{SN}} = {1.1\times 10^{-3}}{}\)
\cite{Higdon_lingenfelter_2005}.
The expected number of SNPs to be born within an OB association follows
from the cluster distribution in
Eq.~(\ref{eq: star cluster star distribution}) multiplying the distribution
by \(f_{\text{SN}}N_*\) and integrating over the desired range. 
In total, the predicted birthrate of SNe is
$2.81\times 10^{4}/$Myr, or about one every 36~years---consistent with the
expected number of Galactic SNe~\cite{Mac_low_1988}.
The cumulative probability distribution \(P(>N_*^{\text{SN}})\) describes
how the SNPs are distributed among the OB associations.
The star formation in an \OBA\ is not continuous, but is concentrated in
several episodes which are typically separated by several
Myrs~\cite{higdon_galactic_2013, MWL1997}.
In the case of $g$  episodes of star formation in an association, the
cumulative probability distribution \(P(>N_*^{\text{SN}})\) is given
by~\cite{Higdon_lingenfelter_2005}
\begin{equation}
 P(>N_*^{\text{SN}}) = C(g) - 0.11 \ln N_*^{\text{SN}}, \quad 1 < N_*^{\text{SN}} \leq 1870 g,
    \label{eq: prob snp content g star forming episodes}
\end{equation}
where \(C(g) = 0.828, 0,95\) and \(1.0\) for \(g = 1, 3\) and \(5\),
respectively. The expected number of SNPs to be born in all Galactic \OBA\
per Myr
is 23270, 26600 and 28100  for $g=\{1,3,5\}$. For five such episodes, this
model predicts that all of the OB stars in our Galaxy would be born in OB
associations. 

For \(g = 1, 3\) and \(5\), the upper limit for the number of massive stars
in OB association becomes, respectively, 1870, 5610 and 9350. How does
this fit with observations and models made by other authors?
Reference~\cite{MWL1997} investigated SNPs with a minimum mass of $8\,M_{\odot}$
and found for their model that the largest associations with five star forming
episodes contained about 7200~OB stars. This is less than what the distribution
of Ref.~\cite{Higdon_lingenfelter_2005} predicts, but Ref.~\cite{MWL1997}
did use a different IMF. Reference~\cite{MWL1997} compares their result
with that of Refs.~\cite{Heiles_1990} and \cite{Ferriere_1995}, and writes
that their numbers correspond to, respectively, a maximum number of SNPs in
Galactic associations of about \(9000\) and \(7230\). Thus, the upper limit
for our model of 9350~SNPs in an association is on the higher end
of this range.

In the right panel of Fig.~\ref{fig: schematic spiral arm structure}, we show
a Monte Carlo realisation of the Milky Way for a simulation time of 100\,Myrs
and five star formation episodes per \OBA.
Each black dot represents an OB~association, containing at least \(2\) OB
stars. The different relative weights of the four main arms as well as the
Local Arm and the devoid region of Sagittarius-Carina are visible.

\subsubsection{Individual SNe}

When the number of SNPs, i.e.\ OB stars with mass $\geq 8M_\odot$, to be born
in OB associations per Myr is specified, a
Monte Carlo simulation is used to draw associations from the distribution
in Eq.~(\ref{eq: star cluster star distribution}) with limits given by
\(N_{\text{min}}^{\text{SN}} = 2\) and \(N_{\text{max}}^{\text{SN}} = 1870\).
Each association is populated \(g\) times from
Eq.~(\ref{eq: star cluster star distribution}), and each star forming episode
in an association is separated by 4\,Myr~\cite{higdon_galactic_2013, MWL1997}.
Reference~\cite{MWL1997} suggests that each star forming episode in an
association results in roughly an equal number of stars, which agrees with
the earlier work of \cite{Elmegreen_lada_1974}. Therefore, we will take the
separate episodes of stars formation to contain an equal number of stars in
our model. Once the associations and their stellar content are drawn, they
are placed in the modelled Galaxy following the probability distribution of
Galactic OB~associations.

Then the masses for the OB stars are drawn randomly from the modified Kroupa
IMF. The mass is then used to determine the lifetime of the star.
Reference~\cite{Schaller_1992} investigated stellar evolutionary models and
also gave
data for the lifetimes of stars as a function of mass. These data were then
used in Ref.~\cite{Fuchs_2006} to fit the main-sequence lifetime in the mass
range \(2 \leq M/M_{\odot} \leq 67\). To make the function
better fit the expected lifetimes for stars with even higher masses
given in Ref.~\cite{Schaller_1992}, we have scaled the function of
\cite{Fuchs_2006} by a factor of \(1.65\):
\begin{equation}
  \tau = 1.65\tau_0 (M/M_\odot)^{-\alpha}  ,\quad \tau_0 = 160\,{\rm Myr},
  \quad \alpha = 0.932.
    \label{eq: age as func of mass}
\end{equation}

The massive OB stars have now been assigned a mass, lifetime and birthplace
in the Galaxy, with the birthplace being the association's center. These
stars are not taken as stationary but will move about as time passes.
Typically, the velocity dispersion within an OB association is of the
order of a few km/s and anisotropic \cite{Wright2020}.
For simplicity, we assume a Gaussian isotropic velocity distribution
with a dispersion of $\sigma = 2$\,km/s.
Finally, we have to account for the possible kick of the surviving star
when the more massive star of a binary system explodes in SN explosions.
We assume that the fraction $f_{\rm bin}=0.7$ of all stars is born in
a bninary system, and use as medium kick velocity
$v_{\rm kick}=8$km/s~\cite{2019A&A...624A..66R}.

\subsection{Known OB associations}\label{ssec:bturb}

We have compiled relevant data and information on observed \OBA s mainly from
the review~\cite{Wright2020}.  We have added data\footnote{For more details see Ref.~\cite{Mi24}.} including the position, age,
and
the total mass of an association in stars with mass $M\geq M_{\min}$ found in
Refs.~\cite{Pecaut_mamajek_2013, Mamajek_2022}. Using then the Kroupa IMF,
the number of stars with mass larger than  $M_{\min}$ is derived
and shown in Table~\ref{table:knownOBA}.

\begin{table}
  \centering
    \begin{tabular}{@{}lrrrrrrrrr@{}}
    Name & $l$/degree & $b$/degree & $d$/pc & $t$/Myr & $N_{\text{s}}$ &  $M_{\min}/M_\odot$ & SNP & SNe & $\leq$\,Myr\\
    Sco-Cen US & 351.50 & 20.00 & 143 & 10 & 107 & \( 1.06\) & 3 & 0& 0 \\
    Sco-Cen UCL & 331.00 & 12.50 & 136 & 16 & 179 & \( 1.06\) & 6 & 1 & 0 \\
    Sco-Cen LCC & 298.50 & 5.50 & 115 & 15 & 147 & \( 1.06\)  & 5 & 1 & 0\\
    Ori OB1a & 201.0 & $-17.3$ & 360 & 12 & 234 & \( 2.0\)  & 22 & 2 & 0\\
    Ori OB1b & 205.0 & $-18.0$ & 400 & 6 & 123 & \( 2.0\)  & 12 & 0 & 0\\
    Ori OB1c & 211.3 & $-19.5$ & 385 & 4 & 246 & \( 2.0\)  & 23 & 0 & 0\\
    Ori OB1d & 209.0 & $-19.5$ & 380 & 1 & 62 & \( 2.0\)  & 6 & 0 & 0\\
    Vela OB2 & 262.80 & $-7.70$ & 400 & 13 & 72 & \( 2.5\)  & 10 & 1 & 0\\
    Trumpler 10 & 262.80 & 0.70 & 372 & 45 & 22 & \( 2.68\)  & 4 & 4 & 0\\
    Cyg A & 72.61 & 2.06 & 1895 & 14 & 125 & \( 2.68\)  & 20 & 3 & 0\\
    Cyg B & 78.58 & 3.31 & 1726 & 9 & 93 & \( 2.68\)  & 15 & 1 & 0\\
    Cyg C & 76.11 & 0.54 & 1713 & 8 & 88 & \( 2.68\)  & 14 & 1 & 0\\
    Cyg D & 75.44 & 1.19 & 2000 & 20 & 86 & \( 2.68\)  & 14 & 4 & 0\\
    Cyg E & 80.19 & 0.85 & 1674 & 8 & 133 & \( 2.68\)  & 21 & 1 & 0\\
    Cyg F & 74.04 & 1.44 & 1985 & 11 & 156 & \( 2.68\)   & 24 & 2 & 0\\
    Cyg OB7 & 89.00 & 0.00 & 630 & 13 & 7 & \( 20\)   & 100 & 13 & 2\\
    Aur 1 & 170.72 & $-0.16$ & 1056 & 21 & 206 & \( 2.68\)  & 33 & 11 & 1\\
    Aur 3 & 170.70 & 0.11 & 1514 & 12 & 111 & \( 2.68\)  & 17 & 2 & 0\\
    Aur 4 & 173.09 & $-0.03$ & 1923 & 2 & 128 & \( 2.68\)  & 20 & 0 & 0\\
    Ara OB1 & 337.68 & $-0.92$ & 1100 & 2 & 11 & \( 20\)   & 54 & 0& 0\\
    Per OB2 & 159.20 & $-17.10$ & 318 & 10 & 9 & \( 2.68\)  & 1 & 0 & 0\\
    Car OB1 & 286.50 & $-0.50$ & 2350 & 6 & 294 & \( 2.68\)  & 45 & 1 & 0\\
    CMa OB1 & 224.00 & $-1.30$ & 1200 & 10 & 93 & \( 2.68\)   & 15 & 1& 0\\
    Mon OB1 & 202.10 & 1.00 & 700 & 5 & 24 & \( 2.68\)  & 4 & 0 & 0\\
    Mon OB2 & 207.35 & $-1.60$ & 1500 & 5 & 13 & \( 20\)   & 69 & 1& 1\\
    Sco OB1 & 343.30 & 1.20 & 1580 & 6 & 28 & \( 20\)   & 157 & 2& 2\\
    Lac OB1 & 96.70 & $-17.60$ & 358 & 10 & 14 & \( 4.7\)  & 6 & 1 & 0\\
    Sct OB2 & 23.18 & $-0.54$ & 1170 & 6 & 14 & \( 2.68\)  & 2 & 0 & 0\\
    Ser OB2 & 18.21 & 1.63 & 1900 & 5 & 107 & \( 2.68\)  & 17 & 0 & 0\\
    Ser OB1 & 16.72 & 0.07 & 1706 & 5 & 20 & \( 20\)  & 106 & 2 & 1\\
    \end{tabular}
    \caption{Observational data on known associations (position $(l,b,d)$, age $t$ and the number of stars $N_s$ with mass $M>M_{\min}$) together with estimated past stellar content (total number of SNP, the number of past SNe, and the number of SNe in the last 1\,Myr).
    \label{table:knownOBA}}
\end{table}

\subsection{Comparison with observational data} \label{ch: modelled and observational OBA}

\begin{figure}
  \centering
  \includegraphics[width=0.85\linewidth]{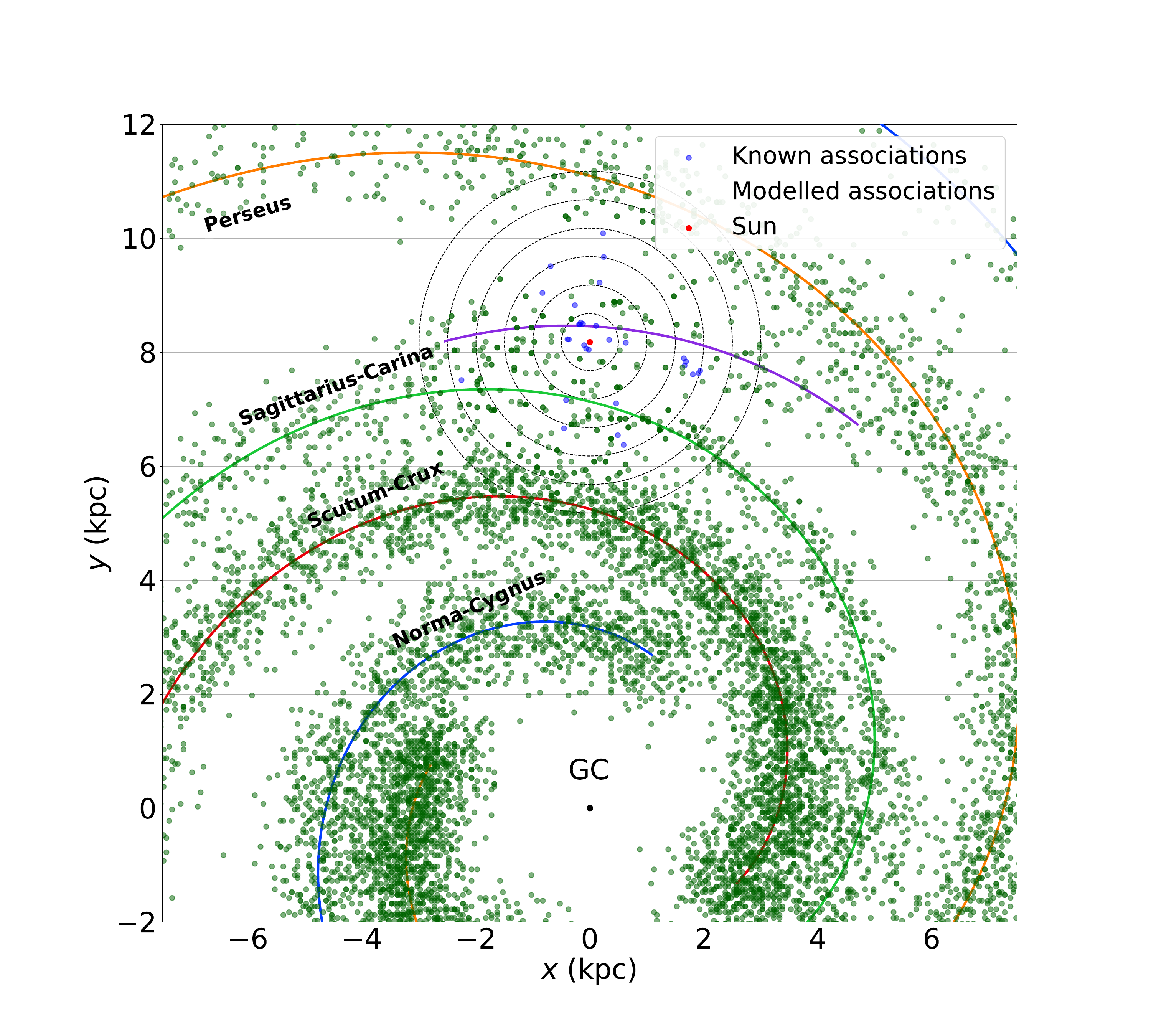}
  \caption{
    Distribution of OB associations after 100\,Myr with five star formation episodes. Blue and green dots represent the known and modelled associations, respectively. The four major spiral arms and the Local Arm passing close to the Sun
    are highlighted by lines.}
    \label{fig: associations_from_galaxy}
\end{figure}

We now combine the modelled and observed associations into one unified model.
The resulting plot of the Galaxy can be seen in right panel of
Fig.~\ref{fig: associations_from_galaxy}. Here, the spiral arms are shown in
the background with labels and the same colours as used in
Fig.~\ref{fig: spiral_arm_nii_intensity_five_arms_gum_cygnus}. The modelled
associations are shown with green circles and the known ones in blue.
The procedure for combining the modelled and known associations will next
be described.

First, we generate a realisation of the Milky Way starting 100\,Myr ago with
five star forming episodes for each association and using the data on the
known associations in Table~\ref{table:knownOBA}. To populate the Galaxy
with associations, we first add the known associations. Then concentric
circles with a spacing of 0.5\,kpc centred on the Sun are generated, which
are also shown in Fig.~\ref{fig: associations_from_galaxy}. For each circular
region, we compare the number of known associations to the number of modelled
associations. If the number of known associations is less than the number
of modelled associations, we randomly draw from the modelled associations
located in this circular region until the sum of associations in this
circular region equals the total number of modelled associations.
Beyond 2.5\,kpc, there are only modelled associations.

As seen from Fig.~\ref{fig: associations_from_galaxy}, this combined model
follows the spiral arm  centerlines, while the devoid region
  of Sagittarius-Carina is barely visible.
This figure explains also why we chose \(\alpha = 0.8\) in
Eq.~(\ref{eq: star cluster star distribution}) instead of \(\alpha = 1\) as
in Ref.~\cite{Higdon_lingenfelter_2005}: One expects that, if not all, we
should have observed the majority of the closest associations to Earth.
Thus, there should be at most only a few modelled associations in the first
circle in Fig.~\ref{fig: associations_from_galaxy},
what we achieve with \(\alpha = 0.8\).
A higher value for \(\alpha\) makes the distribution
in Eq.~(\ref{eq: star cluster star distribution}) steeper, meaning we would
have many more associations with fewer stars, creating an
over-abundance of modelled associations close to Earth. 

\begin{figure}[h!]
    \centering
    \includegraphics[width=1\linewidth]{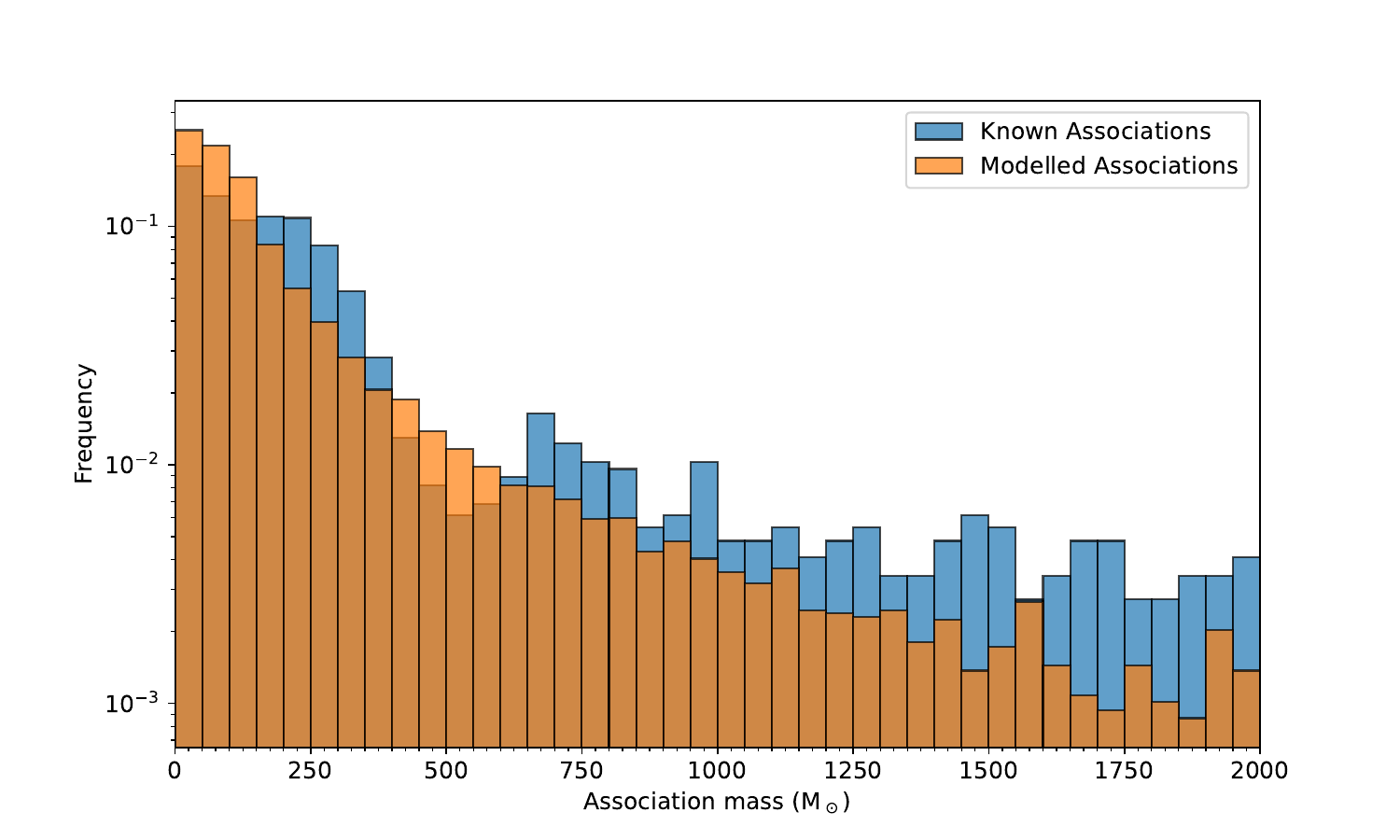}
    \caption{An example of the mass distribution of \OBA s for five star forming episodes over 100\,Myr compared to the one of known \OBA s.
    }
    \label{fig: combined model mass distribution}
\end{figure}

In order to assess how accurate our model for the OB~associations is, we
first estimate the past stellar content of the known OB associations.
To do this, we draw stars from the IMF until the number of surviving stars
agrees with observed number of stars in the given mass range.
In the last three columns of Table \ref{table:knownOBA}, we give the average
number for the total number of stars with masses above \(8 \, M_{\odot}\) after
10,000 iterations of this procedure. We have also given the number of past
SNe that this procedure predicts and how many SNPs are left. Note that
Trumpler 10 is estimated as having no stars left massive enough to produce a
SN due to its old age. 
Finally, we compare the predictions for the mass distribution of stars
in modelled to the one in observed \OBA s,
excluding the contribution of stars with mass $< 8 \, M_{\odot}$.
In Fig.~\ref{fig: combined model mass distribution}, we compare these two
distributions using a bin
width of \(50 \, M_{\odot}\) and five episodes of star formation for the
modelled associations. The exact count in each bin varies for each
run, but overall, the fit between the mass distribution for modelled and
known associations is for \OBA s with $M\lesssim 1000M_\odot$
good, while at higher masses less \OBA s are predicted. This discrepancy
may partially explained by an observational bias favouring the detection of
massive associations. With fewer episodes of star formation, the mass
distribution for the modelled associations gets skewed towards the lower mass
end, and thus, five episodes of star formation results in the best fit. 

\section{Program structure}

The program has been written in \texttt{Python} and has been tested on both
Linux and Windows. The code is  distributed into five main folders:
\texttt{galaxy\_model}, \texttt{nii\_intensities}, \texttt{tests},
\texttt{observational\_data}, and \texttt{utilities}. Most of the figures
in this
article have been generated by the scripts in \texttt{tests}. 

The code generating the Galaxy generates a rather large amount of data, and
the data are therefore written to a file to save RAM. The code automatically
checks if the folders for storing data and figures exist and generates them
if they are
missing, for instance, when running the code for the first time. 

In the folder \texttt{utilities}, the user will find the files
\texttt{constants.py} and \texttt{settings.py}. All parameters for the models are stored within \texttt{constants.py}, and the user can change, for instance, the scale-lengths \texttt{h\_lyc}, \texttt{h\_spiral\_arm} and \texttt{h\_axisymmetric}, and the distance between the Earth and Galactic center, \texttt{r\_s}. Start angles, pitch angles, and maximum and minimum distance from the Galactic center for the four major spiral arms, the Local Arm and the devoid region of the Sagittarius-Carina Arm, are also defined within this file. The parameters for the broken power-law for the IMF, and \(\tau\) and \(\alpha\) which enters Eq.~(\ref{eq: age as func of mass}), are also found here, and the relative paths to where data and figures are stored are defined at the bottom of this file. 

Within \texttt{settings.py}, four variables are defined: \texttt{num\_grid\_subdivisions}, \texttt{add\_local\_arm}, \texttt{add\_devoid\_region\_sagittarius} and \texttt{add\_gum\_cygnus}. The three latter booleans determine whether or not if the Local Arm, the Sagittarius-Carina Arm's devoid region, and the Cygnus X and the Gum Nebula shall be added to the model. These three booleans are by default set to \texttt{True}. \texttt{add\_gum\_cygnus} only affects whether or not Cygnus X and Gum Nebula shall be included in the N\,II intensity plot for the spiral arm model, as these regions already are taken into account from the known data when modelling OB associations. 

The parameter \texttt{num\_grid\_subdivisions} splits up the interpolation
procedure into smaller pieces, with \texttt{num\_grid\_subdivisions} being
the number of pieces. The interpolation procedure is rather memory intensive
due to the large amount of data generated, so to reduce memory needs, the
user can split this workload into several smaller parts. Note that the larger
the value for \texttt{num\_grid\_subdivisions}, the more time is required for
the interpolation to finish. We recommend using \texttt{num\_
  grid\_subdivisions = 1} for \(32 \, \text{GB}\) RAM and \texttt{num\_grid\_subdivisions = 4} for \(8 \, \text{GB}\) RAM.

\texttt{utilities.py} contains various miscellaneous functions used in the calculations. Equation~(\ref{eq: earth_centric to galacto_centric}) is implemented via the functions \texttt{rho(r, l, b)}, \texttt{theta(r, l, b)} and \texttt{z(r, b)}, and converts the Earth-centric coordinates \((r, l, b)\) to Galactocentric coordinates \((\rho, \theta, z)\). The function \texttt{xy\_to\_long(x, y)} takes as input the \(x\) and \(y\) coordinates and calculates the corresponding longitude \(l\). The function \texttt{axisymmetric\_disk\_population(rho, h)}, with \texttt{rho} being the distance from the Galactic center and \texttt{h} the scale-length of the Galactic disk, is the implementation of Eq.~(\ref{eq: axisymmetric_distribution}). Equation (\ref{eq: height_distribution}) is implemented in the function \texttt{height\_distribution(z, sigma)}, with \texttt{z} being the vertical height from the Galactic plane and \texttt{sigma} determining the falloff in density transverse the plane. 

For averaging the N\,II intensities for the longitudinal profiles, we have defined in \texttt{utilities.py} the function \texttt{running\_average(data, window\_size)}. \texttt{data} are the intensities, and \texttt{window\_size} determines the number of points to calculate the average from for each value of \(l\).
The IMF  and the stellar lifetime as a function of initial mass are also
defined here. 

\subsection{\texttt{observational\_data}}

This folder contains programs handling the observational data. For instance,
histograms of H\,II regions can be generated in \texttt{rrl\_hii.py} by
accessing the VizieR database. Part of the data entering
Table~\ref{table:knownOBA} is calculated in \texttt{analysis\_obs\_data.py},
and in \texttt{firas\_data.py} the FIRAS data, which were downloaded
from NASA's Goddard Space Flight Center, \url{https://lambda.gsfc.nasa.gov/product/cobe/firas_prod_table.html}, are prepared for plotting.

\subsection{\texttt{nii\_intensities}}

In this folder, the axisymmetric and spiral arm models are generated by
\texttt{axisymmetric\_disk\_model.py} and \texttt{spiral\_arm\_model.py}.
Cygnus X and Gum Nebula are generated in \texttt{gum\_cygnus.py}, and in
\texttt{chi\_squared.py} is a script optimising the parameters for
the spiral arm model. 

In \texttt{axisymmetric\_disk\_model.py}, we have two functions: \texttt{plot\_axisymmetric()} and \texttt{calc\_modelled\_intensity(b\_max = 5)}, where the former is responsible for loading the generated data and plotting it. The result is shown in Fig.~\ref{fig: axisymmetric_nii_intensity}. For \texttt{calc\_modelled\_intensity(b\_max = 5)} there is one input parameter, namely \texttt{b\_max} whose default value is \(5^\circ\). This is the maximum angle in latitude for which we integrate over such that \(\Delta b = 2 \, \texttt{b\_max}\). This is useful for testing how different limits for the integration over latitude affect the resulting intensity plot. The other parameters entering the model, such as \texttt{rho\_max\_axisymmetric}, can be changed within \texttt{constants.py}. The resulting intensities for the axisymmetric model are saved to a \texttt{.npy} file.

In \texttt{spiral\_arm\_model.py}, the function \texttt{calc\_modelled\_intensity()} calculates the spiral arm intensities.  This function has the following
eleven input parameters:
\begin{itemize}
    \item \texttt{readfile\_effective\_area=True}: This boolean value controls whether or not the effective area per spiral arm shall be calculated. If it is set to \texttt{True} the values will be read from file, and if it is set to \texttt{False} the effective areas will be calculated and saved to file. It should be set to \texttt{False} when the code is run for the first time and whenever parameters for the spiral arms are changed, as this will affect the resulting effective areas. 

    \item \texttt{interpolate\_all\_arms=True}: This boolean controls whether or not the densities for each arm shall be interpolated. If set to \texttt{True}, each spiral arm is interpolated over the entire Galactic plane, and then saved to file. The default value is set to \texttt{True}, as it is assumed the user will call the function \texttt{calc\_modelled\_intensity} only when spiral arm parameters are changed and thus need to recalculate the intensities. The reason for having this boolean in the first place is due to how the script for optimising the spiral arm parameters in \texttt{chi\_squared.py} is designed. In short, we change the parameters for one arm at a time while keeping the others the same, and thus, we interpolate only one arm at a time. As such, we avoid interpolating the other arms multiple times for the same parameters, saving time. The intensities still need to be calculated, so we call \texttt{calc\_modelled\_intensity} with \texttt{interpolate\_all\_arms=False}. It should be set to \texttt{True} for all other uses.

    \item \texttt{calc\_gum\_cyg=False}: This boolean controls whether or not the modelled N\,II intensities from Cygnus X and Gum Nebula should be calculated. If set to \texttt{True}, the values will be calculated and saved to file by the script in \texttt{gum\_cygnus.py}. It should be set to \texttt{True} the first time the script is run, and each time parameters for either Cygnus X or Gum Nebula are changed. Otherwise, it should be set to \texttt{False} to save time. 

    \item \texttt{recalculate\_coordinates=False}: The fourth and final boolean for this function controls whether or not the Galaxy coordinates should be recalculated. This process is time-consuming and should only be set to \texttt{True} the first time the script is run or if parameters for the Galaxy are changed, such as the maximum extent for the spiral arm \texttt{rho\_max}, which is located in \texttt{constants.py}.

    \item \texttt{b\_max=5} and \texttt{db\_above\_1\_deg=0.2}: These parameters define the integration range over latitude and the step-size above $1^\circ$ latitude. The lines-of-sight are sampled non-uniformly for \(|b| \leq 1^\circ\) and with a fixed \(\d b\) for \(|b| > 1^\circ\) up to \texttt{b\_max}. 

    \item \texttt{fractional\_contribution}, \texttt{h}, \texttt{sigma\_arm}, \texttt{arm\_angles} and \texttt{pitch\_angles} are the final parameters for this function, and are the parameters for the arms themselves as well as the density scale length \texttt{h} which enters Equation (\ref{eq: arm median density}). By default, these values are set to those stored in \texttt{constants.py}, but could be set to different values when calling the function. \texttt{sigma\_arm} corresponds to \(\sigma_{\text{A}}\) which enters Equation (\ref{eq: transverse density spiral arms}). \texttt{fractional\_contribution}, \texttt{arm\_angles} and \texttt{pitch\_angles} have to contain the same number of values and could either be a \texttt{Python list} or a \texttt{numpy array}.
\end{itemize}

As briefly mentioned, \texttt{chi\_squared.py} contains code which optimises the parameters entering the model. To run this script, one only has to call the function \texttt{run\_tests(num\_iterations=10)} and specify how many iterations you want the code to run. The default value is set to \(10\). For the first iteration, every parameter is varied around the value located in \texttt{constants.py}. The best parameters are kept and used as input in the next iteration. The best-fit parameters are printed to the terminal when the script is finished.

In \texttt{gum\_cygnus.py}, we generate and save to file the intensities resulting from the modelled Cygnus X region and Gum Nebula. The calculations for both are done by calling the function \texttt{generate\_gum\_cygnus()}, which in turn calls the functions \texttt{gum()} and \texttt{cygnus()}. The latter functions calculate the Gum Nebula and the Cygnus X region, respectively, as described in the text. If the user wants to change the parameters for these regions, this can be done within \texttt{gum()} and \texttt{cygnus()}.

\subsection{\texttt{galaxy\_model}}

The final folder contains the scripts simulating the OB associations. The
first file, \texttt{galaxy\_density\_distr.py}, generates the coordinates
for the Galaxy. The grid is generated in  \((x, y, z)\) coordinates with uniform spacing between each point. The generated coordinates and densities are saved to file and are generated by calling the function \texttt{generate\_coords\_densities()}. The parameters which enter this function are:
\begin{itemize}
    \item \texttt{plane=1000}: This is the number of points along the \(x\) and \(y\) direction whose default value is \(1000\), such that the Galactic plane has a total of \(10^6\) number of points at which the OB associations can be placed.

    \item \texttt{transverse=20}: This is the number of points along the \(z\) direction whose default values is \(20\). As such, with the default value for \texttt{plane}, the number of grid points for which the OB associations could be placed is \(20\) million. 

    \item \texttt{half\_edge=40}: This is the distance in kpc from the Galactic center along the \(x\) and \(y\) axis for which the Galaxy is defined. The default value is set to 40\,kpc, which is the distance from the Galactic center to the endpoint of the spiral arms 35\,kpc plus 5\,kpc to take into account the falloff in density transverse the spiral arms. 

    \item \texttt{readfile\_effective\_area=True}: This boolean parameter does the same job as in \texttt{calc\_modelled\_intensity}. Set this to \texttt{False} the first time the code is run and anytime parameters for the modelled Galaxy are changed. Otherwise, set it to \texttt{True} to save time running the code. 

    \item \texttt{read\_data\_from\_file=True}: This boolean parameter, which defaults to \texttt{True}, controls whether or not the coordinates and densities are read from file or calculated. Set this parameter to \texttt{False} the first time the code is run to calculate coordinates and densities and save them to file. When it is set to \texttt{True}, coordinates and densities are read from file.
\end{itemize}

The code uses an object-oriented approach for modelling OB associations. We have three main classes: one for the Galaxy, one for the OB associations, and one for the OB stars. These are, respectively, located in the following files: \texttt{galaxy\_class.py}, \texttt{association\_class.py} and \texttt{supernovae\_class.py}. 
The Galaxy class is mainly responsible for generating and storing the Galaxy's coordinates and densities, and drawing the positions of OB associations with a Monte Carlo simulation every Myr. When creating an instance of the Galaxy class, three parameters can be passed to it: \texttt{star\_formation\_episodes}, \texttt{sim\_time\_duration} and \texttt{read\_data\_from\_file}. \texttt{sim\_time\_duration} is the number of Myr ago the simulation shall start. The Galaxy is generated in the past to produce SNe which could have produced cosmic rays observable today. Every Myr, we draw the association's placement from the emissivity map generated by \texttt{generate\_coords\_densities()}. The parameter \texttt{read\_data\_from\_file} is passed on to \texttt{generate\_coords\_densities()}, and as mentioned, controls whether or not the coordinates and densities for the Galaxy are calculated or read from file. 

The final parameter for the Galaxy class is \texttt{star\_formation\_episodes}, which controls how many star forming episodes shall occur in each association.
Each episode is separated by \(4\) Myr and they are equal in size.
Finally, the parameter \(\alpha\), which enters Eq.~(\ref{eq: star cluster star distribution}), can be changed using the \texttt{alpha} argument of the function \texttt{\_association\_distribution} located within the Galaxy class. 

The Galaxy Class generates associations for every Myr by creating instances of the association class, which in turn creates instances of the supernovae class to store information about each OB star. If the user wants to modify the calculated age or the IMF used, this is done in \texttt{utilities.py}.

\section{Summary}

We have presented a model for the longitudinal profile of the N\,II intensity
in the Galactic plane. The model is based on four logarithmic spiral arms,
to which features like the Local Arm and local sources are added.
Since N\,II is a tracer of the line emission of OB~associations, this
model allows one to generate the spatial and temporal average distribution
of  OB~associations and, thence, of core-collapse supernovae in the Milky Way.
In addition to this average population, the model includes supernovae from
known OB~associations, providing thereby a more accurate description of the
nearby Galaxy.

The complete model is made publicly available in the python code {\tt SNOB}. 
The code can be used to generate the source distribution of Galactic CRs,
and the distribution of birth places of pulsars or black holes. Moreover,
accounting for H\,II regions might be useful in the interpretation of radio
dispersion and scattering measurements as electron density
tracers.

\section*{Acknowledgements}
\noindent
We would like to thank D.J.~Fixsen for helpful advice on the FIRAS data, and
D.~Misra and D.~Semikoz for useful suggestions.


\end{document}